# A Serious Game for Introducing Software Engineering Ethics to University Students

Michalis Xenos, Vasiliki Velli

Computer Engineering and Informatics Department, Patras University
xenos@ceid.upatras.gr, velli@ceid.upatras.gr

**Abstract.** This paper presents a game based on storytelling, in which the players are faced with ethical dilemmas related to software engineering specific issues. The players' choices have consequences on how the story unfolds and could lead to various alternative endings. This Ethics Game was used as a tool to mediate the learning activity and it was evaluated by 144 students during a Software Engineering Course on the 2017-2018 academic year. This evaluation was based on a within-subject pre-post design methodology and provided insights on the students learning gain (academic performance), as well as on the students' perceived educational experience. In addition, it provided the results of the students' usability evaluation of the Ethics Game. The results indicated that the students did improve their knowledge about software engineering ethics by playing this game. Also, they considered this game to be a useful educational tool and of high usability. Female students had statistically significant higher knowledge gain and higher evaluation scores than male students, while no statistically significant differences were measured in groups based on the year of study.

## 1　Introduction

While introducing students into software engineering ethics has been recognized as a necessity and many undergraduate computer science and computer engineering programs offer relative courses [1, 2], these courses are mostly based on lessons learned and theoretical essays [3, 4]. We argue that the use of game-based learning (or serious games [5]) into the area of software ethics could be a very helpful practice for the students. This practice allows the students to experience real life situations, during playing and while having fun. These situations are presented in scenarios that require moral judgement and solving ethical dilemmas. To the best of our knowledge, while there are serious games available for personal and social ethics [6], this is the first time a game is used for teaching software engineering ethics, that is based on students' choices on various scenarios that simulate real life situations.

The rest of the paper is structured as follows: In section 2 we present the Ethics Game, the tool used to mediate the learning activity of software engineering ethics. In section 3 we present the setting and the materials of an evaluation study that involved 144 students and in section 4 we present the results of this study. Finally, in section 5 we discuss conclusions and future work.

## 2 The Ethics Game

A game based on a commercial storytelling platform[1] was developed, so to be available in most mobile devices. After installing the application on their mobile or tablet, anyone interested to play the Ethics Game have the option to either search for "*Ethical_Dilemmas*" (this was the title the game was registered) or to download the game using a direct link[2] offered by us. The game introduces the players to basic ethical dilemmas related to software engineering, based on the software engineering code of ethics [7], and offers them choices that will influence how the story will unfold. Some selections are straightforward and force the story to evolve in different paths, and in some cases in alternative endings, while other selections add up to internal scores (not revealed to the user) that also direct the path the story will follow.

The platform application is free to download and offers a variety of commercial stories/games in various genres (romance, drama, Hollywood, fantasy, mystery, comedy, action/adventure, and thriller/horror). Players selecting these stories need to purchase "diamonds" (the platform currency) to unfold the stories, but the rest of the stories/games created by users (as in our case) are free. Most of the commercial stories available on this platform seem to address female players (most of the main characters are females), so one of our research goals was to investigate if this was story-dependent, or is it apply in our Ethics Game as well.

---

[1] https://www.episodeinteractive.com/
[2] https://www.episodeinteractive.com/s/6429627363229696

To play each episode the player is using a "pass". Each user is given 3 passes at the first time they download the application and passes are frequently replenished, since the application adds 3 more passes every 4 hours. A user can also purchase passes, but this is something optional and we advised students against it. Having the passes limitation in mind, we have created a story that unfolds in one week in work. During this week, Rose a software engineer working in a large firm, is facing with a lot of ethical dilemmas. The Ethics Game lasts for 6 days (corresponding to 6 episodes), therefore one needs only 6 passes to finish it. This ensures that every player can start playing the first 3 days of the game, then wait for 4 hours to get more passes and then finish the game. Each one of the first 5 episodes represent one day at work (Monday to Friday), while the 6[th] episode presents one of the alternative endings, based on the player's choices. Fig. 1 presents three instances from the Ethics Game; in the left and right images, the heroine Rose, is about to make a choice facing two alternative options, while in the image in the middle the game designer had zoomed on Rose to emphasise her comments.

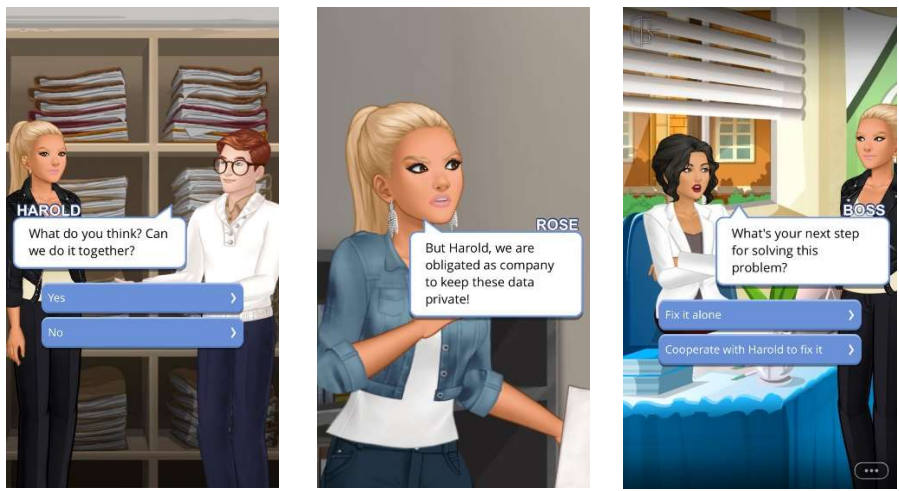

**Fig. 1.** Screenshots from the Ethics Game

The author/creator of a story/game uses commands to change the background and to introduce animated backgrounds and sound. They can direct the story using commands such as (all the following commands are from the Ethics Game): `@ROSE changes into work_outfit_4, @HAROLD stands screen right AND HAROLD faces left, @zoom on ROSE to 200% in 1.5, ROSE (talk_argue_defensive)`. The author can control the flow of the story (how the story will unfold) based on the choices the player selected, using a `choice()` command. Different choices lead to different branches of the story, which are controlled with `if…then…else` commands. Finally, the author can use variables to keep score of various elements related to the story. In our case we used choices and variables to evaluate the players' performance and to present them with alternative endings, based on their choices within the game.

Three alternative endings are available in the Ethics Game. Should a player manage to face all challenges successfully, the game ends with Rose being promoted. Players that made a lot of incorrect choices and repeatedly violated software engineering ethics they end the game by being fired from their position, while the rest of the players remain at their position, but at the end of the game they receive information about what they should have done better.

## 3  The evaluation study

This paper reports a within-subject pre-post study that investigates the learning effectiveness of the Ethics Game (post-test) compared to the lecture-based instruction (pre-test) in the context of campus-based higher education. In specific, these research questions were investigated by this study:

- RQ1: Is there any effect of the Ethics Game activity on students' learning performance?
- RQ2: Did students find the Ethics Game a useful educational tool?
- RQ3: Did students find the Ethics Game a usable tool?
- RQ4: Are there any differences in learning gain, perceived usability and perceived educational effectiveness related to gender?
- RQ5: Are there any differences in learning gain, perceived usability and perceived educational effectiveness related to the year of studies?

The study took place in the context of campus-based classroom education and in specific in the course named "*CEID_Y232: Software Engineering*", during the academic year 2017-2018. This is a required course, offered to the students of the Computer Engineering and Informatics Department (CEID) at the University of Patras, during the second semester of their 4<sup>th</sup> year of studies (8<sup>th</sup> semester). CEID is a 5-year B.Sc. degree with an Integrated M.Sc., corresponding to 300 European Credit Transfer System (ECTS) units. The CEID_Y232 course includes 13 lectures, 4 compulsory assignments and 4 short elective assignments and offers 6 ECTS units to the students. All assignments are graded, the compulsory ones contribute to 30% of the final course grade, while a passing grade in all assignments is a prerequisite for participating in the final exams for the other 70% of the course grade. Participation in the short elective assignments is offered to aid students to improve the overall course grade. Students participating in CEID_Y232 course are introduced to basic ethical dilemmas related to software engineering during the 10<sup>th</sup> lecture and had the opportunity to play the game as part of the 4<sup>th</sup> elective assignment right after the lecture.

### 3.1 Participants

For the examined academic year 2017-2018, 378 students were registered for this course, but only 217 of them submitted all the required assignments so to be able to participate in the exams. Using the Ethics Game to mediate the learning activity was an elective assignment that contributed only 3% of their final grade, but nevertheless a total of 144 students have successfully completed it. This was the only one elective assignment having such a large participation, regardless of the small gain in the final grade, probably since playing a game was considered fun. The other three elective assignments had only: 32, 17 and 11 participants.

Before offering the Ethics Game to the students, a pilot test of the entire procedure performed by using 5 participants, whose responses were excluded from the results. Following the pilot test, 144 students participated in this study, 29 of them were female (20.1%) which is typical in computer engineering studies in Greece. From these students, 63 of them were at the 4<sup>th</sup> year of studies (current year), while 81 students belonged to higher years (students that had failed the CEID_Y232 course and repeating it). Most of the students used their smartphone to play (N=134, 93.1%), while very few used a tablet (N=10, 6.9%) or an emulator on their laptop (N=1, 0.7%).

### 3.2 Materials

The students were asked to complete a knowledge test (pre-test) after the end of the lecture and before downloading the game. They had access to the game only after completing the knowledge test. The knowledge test included all issues related to software engineering ethics that were addressed in the game and comprised of 10 multiple-choice questions with four possible answers each. Then the students were asked to play all six episodes of the Ethics Game and report their result within the game, when they reach one of the possible endings. Finally, after finishing the game, the same test (post-test) was offered to them, and completing it was required to formally finalise this assignment.

Additionally, they were also asked to complete three additional scales as part of the post-test: a) a 3-items scale rating their educational experience with the game from 1 to 5, b) the standardized System Usability Scale [8], provided in participants' native language [9], and c) the 7-point adjective rating question [10] with wordings from "worst-imaginable" to "best-imaginable". Finally, they had the option to comment, on an open question, on issues they feel we could improve in a future version of the game. This is a typical set of scales we have successfully used in similar usability evaluation studies [11-13]. The collected data were organized and pre-processed using Microsoft Excel 365 ProPlus and were analysed using IBM SPSS Statistics v20.0.

### 4 Results

First, reliability analysis was conducted for the questionnaires used in the study. To this end, the Cronbach's alpha measure of internal consistency was used [14]. The 10 questions knowledge test has marginal reliability, (Cronbach's alpha=0.699, N=10). We measured that removing the question number 9 could improve at (Cronbach's alpha=0.715, N=9), while removing any other question would resulted a Cronbach's alpha below the 0.700 threshold, but since the internal consistency was very close to the accepted limit for the 10 questions we decided to keep our original set of 10 questions for compliance with our educational model. SUS is a standardized scale [9, 15, 16] and had also adequate reliability for our dataset (Cronbach's alpha=0.717, N=10). Finally, the educational experience scale had also adequate internal consistency for our dataset (Cronbach's alpha=0.746, N=3).

Following the rationale reported in [17], we produced a composite variable for the normalized learning gain defined as the difference between post-test score and pre-test score ("observed gain") divided by the difference between the max possible score and the pre-test score ("amount of possible learning that could be achieved" [17]).

Table **1** presents descriptive statistics of the dependent variables measured in this study (mean, median, standard deviation and 95% confidence interval).

**Table 1.** Descriptive statistics of the depended variables of this study. Sample size N=144.

| Variable | M | Mdn | SD | 95% CI |
|---|---|---|---|---|
| Pre-test score (0-100) | 67.99 | 70.00 | 16.92 | [65.20, 70.77] |
| Post-test score (0-100) | 80.59 | 85.00 | 15.57 | [78.03, 83.16] |
| Normalized learning gain (%) | 35.04 | 33.00 | 46.01 | [27.46, 42.62] |
| Educational experience rating (1–5) | 3.87 | 4.00 | 0.80 | [3.74, 4.00] |
| SUS score | 83.14 | 85.00 | 8.06 | [81.81, 84.47] |
| Usability adjective rating | 4.84 | 5.00 | 0.86 | [4.70, 4.98] |

### 4.1 Effect on students' learning performance

For the 144 students, only 14 had negative ranks between the pre-test and the post-test, 27 had the exact same score and 103 had positive ranks. This is also reported from the mean score of the composite variable "observed gain" which was high (+35.04%). Additionally, because the data for both test variables were skewed, a Wilcoxon Signed-Ranks test was run, and the output indicated that post-test scores were statistically significantly higher than pre-test scores (Z=8.321, p<.000). Therefore, for RQ1, we can argue that the students did learn about software engineering ethics by playing the Ethics Game.

**Table 2.** Descriptive statistics of students' self-reported ratings of their educational experience

| Question (1: strongly disagree; 5: strongly agree) | M | Mdn | SD | 95% CI |
|---|---|---|---|---|
|  |  |  |  |  |
| Q1: I think that the Ethics Game is useful as an Educational tool | 3.63 | 4.00 | 0.95 | [3.47, 3.78] |
| Q2: I would recommend the Ethics Game to a colleague or friend who wants to learn about ethics in software engineering | 3.80 | 4.00 | 1.11 | [3.61, 3.98] |
| Q3: I would recommend the Ethics Game to a colleague or friend who wants to design something similar | 4.19 | 4.00 | 0.93 | [4.04, 4.35] |
| Overall scale (Cronbach's alpha = 0.746) | 3.87 | 4.00 | 0.80 | [3.74, 4.00] |

### 4.2 The Ethics Game as a useful educational tool

Participating students rated their learning experience with the Ethics Game in the post-test questionnaire. Table 2 presents descriptive statistics of these ratings per question and overall. So, regarding RQ2, the students self-reported ratings for their educational experience for the Ethics Game were relatively high (M=3.87, SD=0.8). In addition, students provided rather positive comments for their educational experience while playing the Ethics Game in the open-ended question of the post-test questionnaire.

### 4.3 The Ethics Game perceived usability

To assess the RQ3, after playing the Ethics Game, the participating students completed the SUS questionnaire and the adjective rating scale, both measures of a system's perceived usability. The Ethics Game received a mean SUS score of 83.14 (SD=8.06). According to a dataset of nearly 1000 SUS surveys [10], this means that students found KLM-FA as "Good to Excellent" (SUS score from 71.4 to 85.5) in terms of perceived usability. Students' usability adjective ratings were also rather high (M = 4.84, SD = 0.86), indicating that the Ethics Game was perceived as "Good" (corresponds to 5).

Table 3. Descriptive statistics of the depended variables of this study, grouped by gender

| Group | Variable | M | Mdn | SD | 95% CI |
|---|---|---|---|---|---|
| Male students (N=115) | Normalized learning gain (%) | 20.42 | 28.50 | 53.83 | [-0.05, 40.90] |
| | Educational experience rating (1–5) | 3.37 | 3.33 | 0.98 | [2.99, 3.74] |
|  | SUS score | 79.05 | 80.00 | 8.00 | [76.01, 82.10] |
|  | Usability adjective rating | 4.24 | 4.00 | 0.99 | [3.87, 4.61] |
| Female students (N=29) | Normalized learning gain (%) | 53.60 | 71.20 | 45.26 | [36.38, 70.82] |
|  | Educational experience rating (1–5) | 4.24 | 4.33 | 0.51 | [4.05, 4.44] |
|  | SUS score | 85.34 | 87.50 | 7.70 | [82.42, 88.27] |
|  | Usability adjective rating | 5.34 | 5.00 | 0.72 | [5.07, 5.62] |

### 4.4 Gender differences

As shown in Table 3, female students evaluated the Ethics Game higher than the male students in both SUS and usability adjective rating scores. Additionally, their self-reported ratings for their educational experience was also higher compared to male students and they also had higher normalized learning gain.

Since the assumption of normality was violated for all groups as measured by the Kolmogorov-Smirnov test, except for the SUS score for male students, which was also marginal (Sig=0.200), a non-parametric test, the two-tailed Man-Whitney U test, was selected. The Normalized learning gain was higher for female students (M=53.60, SD=45.26) than male students (M=20.42, SD=53.83) and using a two-tailed Man-Whitney U test, we have found that the normalized learning gain was statistically significant higher for the female students (U=1172, Z=-2.477, p=0.013).

Regarding the SUS, although the female students evaluated the Ethics Game higher (M=85.34, SD=7.70) compared to male students (M=79.05, SD=8.00), the two-tailed Man-Whitney U test showed that these differences were not statistically significant (U=1282.5, Z=-1.929, p=0.054), although the result was marginal. For the usability adjective rating the female students evaluated the Ethics Game higher (M=5.34, SD=0.72) compared to male students (M=4.24, SD=0.99) and the two-tailed Man-Whitney U test revealed that the adjective evaluation was statistically significant higher for the female students (U=1016.5, Z=-3.566, p=0.000). Finally, for the self-reported ratings for their educational experience female students had also higher reported rankings (M=4.24, SD=0.51) compared to male students (M=3.37, SD=0.98) and the two-tailed Man-Whitney U test confirmed this (U=1086, Z=-2.925, p=0.003).

In conclude, for the RQ4, we can argue that female students enjoyed the Ethics Game more than male students, they found it a more valuable educational tool than the male students did, and by playing the game they improved their knowledge on the field significantly more than what male students did.

### 4.5 Differences related to the year of studies

Table 4 presents the descriptive statistics for the groups of students that were at the current (4[th]) year or being in higher years. This could also be measured by grouping by age, since in our case was equivalent. Since the assumption of normality was violated for all groups as measured by the Kolmogorov-Smirnov test, a non-parametric test, the two-tailed Man-Whitney U test, was selected. Using this test, on the one hand, no statistically significant differences were found for these two groups for the normalized learning gain (U=2470, Z=-0.329, p=0.742) as well as for the self-reported educational experience (U=2153.5, Z=-1.619, p=0.106) and for the usability adjective rating (U=2159.5, Z=-1.736, p=0.083). On the other hand, the usability evaluation of the Ethics Game, based on the SUS, was statistically significantly higher for the students of the current year than this of students from higher years (U=1932, Z=-2.509, p=0.012). Overall, regarding the RQ5, there are no statistically significant differences for students in different year of studies, with the exception that students on the current year evaluated significantly higher the Ethics Game usability based on the SUS.

**Table 4.** Descriptive statistics of the depended variables of this study, grouped by academic year

| Group | Variable | M | Mdn | SD | 95% CI |
|---|---|---|---|---|---|
| Current year students (N=63) | Normalized learning gain (%) | 34.94 | 33.10 | 43.66 | [23.95, 45.94] |
| | Educational experience rating (1–5) | 5.00 | 5.00 | 0.70 | [4.82, 5.18] |
| | SUS score | 85.16 | 85.00 | 6.67 | [83.48, 86.84] |
| | Usability adjective rating | 4.00 | 4.00 | 0.73 | [3.82, 4.18] |
| Students from higher years (N=81) | Normalized learning gain (%) | 33.50 | 33.30 | 50.75 | [20.72, 46.29] |
| | Educational experience rating (1–5) | 3.66 | 3.67 | 0.87 | [3.44, 3.88] |
| | SUS score | 80.87 | 82.50 | 8.49 | [78.74, 83.01] |

| | Usability adjective rating | 4.60 | 5.00 | 0.96 | [4.36, 4.84] |

## 5 Conclusion and Future Goals

Although the game wasn't aimed to be particularly difficult, from the 144 students that participated, only 37 (25.7%) managed to reach the end where Rose gets promoted. At least, only 8 students (5.6%) reached the end where she gets fired. The rest of the 99 students (68.8%) reached the end where Rose remained at her position and they could read about which ethic related issues should be more careful at. This was one of the things that most students comment on, asking to remove it from an updated version. They mention that this was like spoilers that prevent them from playing again and they would prefer to play again and find out by themselves if they are able to perform better. This, among other issues, was improved in the version available today in the platform. Another issue that mentioned by the students is that they didn't like the waiting time for their passes to be refilled, so they proposed to combine days into three larger episodes, so they could play them all in once.

Since we have asked the students to fill-in the post-test questionnaire just after they had finished the game, we don't have any accurate measure of how many students kept playing the Ethics Game, even after fulfilling the requirement of the elective assignment. We can see that the Ethics Game had 921 reads, 14 days after the deadline of the corresponding assignment, which means that many of them probably did play it a few more times, but the platform does not report back on individual number of plays. This is one of the future goals to investigate further. Adding more episodes and expanding the game on other issues related to software engineering is another future goal.

## Acknowledgment

The authors would like to thank the 144 students that participated in this study and helped us with their comments.This is a pre-print version of the paper. Please cite this paper as:

Michalis, X., & Vasiliki, V. (2018). *A Serious Game for Introducing Software Engineering Ethics to University Students*. 21st International Conference on Interactive Collaborative Learning, ICL2018, Kos, Greece, pp. 263-274, September, 2018.